\documentclass[12pt]{article}
\usepackage{graphicx}
\usepackage{subfigure}
\usepackage{amsmath}
\usepackage{amssymb}

\usepackage{lipsum}
\usepackage{blindtext}

\makeatletter
\newenvironment{tablehere}
{\def\@captype{table}}
{}

\makeatother
\begin{document}
\begin{center}
\begin{large}
\title\\{ \textbf{Masses and Decay constants  of Heavy Flavour mesons in perturbative approach}}\\\
\end{large}

\author\

\textbf{$ Jugal\;Lahkar^{\emph{a,b}}\footnotemark \:\:,\: D\:K\:Choudhury^{\emph{a,c}}and\: B\:J\:Hazarika^{\emph{c}}$ } \\

\footnotetext{Corresponding author. e-mail :  \emph{neetju77@gmail.com}}
\textbf{a}. Dept.of Physics, Gauhati University, Guwahati-781014, India.\\
\textbf{a}. Dept.of Physics, Tezpur University, Tezpur-784028, India.\\
\textbf{c}. Centre of theoretical Physics, Pandu College, Guwahati-781012, India.\\

\begin{abstract}
Determination of wave function is very essential in the calculation of static and dynamic properties like masses
and decay constants of pseudo scalar mesons and masses of vector mesons . We useDalgarno's perturbation theory and Variationally Improved Perturbation Theory (VIPT) to solve the
Schroedinger equation in a QCD inspired potential model with the Cornell potential which consists of coulomb and
linear potential.
We use both the options like Columbic parent firstly and then linear parent while using the VIPT method.
Comparison of both these methods and also with experimental results are done in the calculation of masses and decay constants.

\end{abstract}
\end{center}
\

\section{Introduction:}\rm
In recent years,various approximation methods[1] have been used to explore the static and dynamic properties of heavy flavour mesons in non-perturbative QCD inspired potential model where solution of Schrodinger equation is essential.  These are mainly:\\
$1.$ Dalgarno,s perturbation theory[2-6],\\
$2.$ Variational Method[7-9],\\
$3.$ WKB Approximation[16],\\
$4.$ Variationally Improved Perturbation Theory[10-15].\\\\\
  Out of these four,the last one is comparatively new which combines both the variational method and the stationary state perturbation theory and is reported to solve time independent problems of quantum mechanics. The method was put forwarded by You S.K.et al[11] and later applied to heavy quarks by Aitchinson and Dudek[12] for the linear plus coulomb potential($-\frac{4\alpha_s}{3r}+br$).  The method is now well known as : $Variationally Improved Perturbation Theory(VIPT)$.  Later,it was applied by Fernandez[13] in the calculation of exact perturbation correction to energy and wave-function for the same potential.\\\\
  
The method substantially reduces the limitations of usual perturbation theory through the use of Variational method over it.  In usual perturbation theory,the parent potential must be strong enough compared to the perturbed part so that the results can be expressed in a converging finite series upto higher orders.  However,the smallness of the parameters in the chosen parent potential the convergence may be lost and then we use the variational method with a known trial wave function  and then optimize it to get the new parameters which would make the parent potential strong enough over the perturbed one to use perturbation theory(eg.$\alpha'$ instead of $\alpha$,with $\alpha'>\alpha$).One advantage of VIPT is that it does not concern about whether we have a good unperturbed Hamiltonian or not.\\\\

    With these motivation,in the present work,we have extended this method(VIPT) to heavy-light mesons to estimate some of their static properties,such as:Masses,Leptonic Decay Constants,Oscillation frequency etc.  We recall that while using VIPT to heavy quark Physics for linear plus coulomb Cornell potential, we have two choices:\\\
   $1.$Coulomb as Parent and Linear as Perturbation,\\
   $2.$Linear as Parent and Coulomb as Perturbation.\\\
 We consider both the options in our calculations. A brief comparison is made between the two options,as well as with the results obtained from Dalgarno's perturbation theory[17] and our recent results with only variational approach[18]. At the same instant,we also compare our results with those of more advanced tools like lattice QCD and QCD Sum Rules.
\\\\
The manuscript is arranged as:in the section 2 we give the formalism with two methods:Dalgarno's perturbation theory and Variationally Improved Perturbation Theory,in section 3 we discuss the results on masses and decay constants of Heavy Flavour mesons and in section 4 we summarize the conclusion.\\
\section{Formalism:}\rm
\subsection{Dalgarno's perturbation Theory:}
In Dalgarno's perturbation theory, we make small deformation to the Hamiltonian of the system,

\begin{equation}
\label{H}
H=H_0+H^\prime,
\end{equation}

where $H_0$ is the Hamiltonian of the unperturbed system and $H^\prime$ is the perturbed Hamiltonian. The approximation method is most suitable when $H$ is close to the unperturbed Hamiltonian $H_0$, i.e. $H^\prime$ is small. \\
The standard potential is[19],

\begin{equation}
\label{potentialc0}
V(r)=-\frac{4\alpha_s}{3r}+br. 
\end{equation}

This Coulomb-plus-linear potential, called Cornell potential is an important ingredient of the model which is established on the two kinds of asymptotic behaviours: ultraviolet at short distance (Coulomb like) and infrared at large distance (linear confinement term).\\

The Schr$\ddot{o}$dinger equation takes the form

\begin{equation}
H|\psi\rangle=(H_0+H^\prime)|\psi\rangle=E|\psi\rangle,
\end{equation}

so that the first-order perturbed eigenfunction $\psi^{(1)}$ and eigen energy $W^{(1)}$ can be obtained using the relation

\begin{equation}	
\label{sch2}
H_0 \psi^{(1)} + H^\prime \psi^{(0)}=W^{(0)}\psi^{(1)} + W^{(1)} \psi^{(0)},
\end{equation}

where $H_0$ is the parent Hamiltonian defined as,

\begin{equation}
H_0=\frac{-\bigtriangledown^2}{2\mu}+V(r)
\end{equation}

and

\begin{equation}
\label{W0}
W^{(0)}=  <\psi^{(0)}\vert H_0 \vert \psi^{(0)}>,
\end{equation}

\begin{equation}
\label{W1}
W^{(1)}= <\psi^{(0)}\vert H^\prime \vert \psi^{(0)}>.
\end{equation}
With Cornell potential we get two choices:\\\\
$1.$Coulomb Parent Linear Perturbation.\\
$2.$Linear Parent Coulomb Perturbation.\\\\

$1.$\underline{Coulomb Parent Linear Perturbation}:\\\\
For the first option using Dalgarno's perturbation theory,we get the total wave-function as,
\begin{equation}
\label{psic}
\psi^{total}_I(r)=\psi^{(0)}_I(r)+\psi^{(1)}_I(r)=\frac{N}{\sqrt{\pi a_0^3}}\left[ 1-\frac{1}{2}\mu b a_0r^2\right] e^{-\frac{r}{a_0}}, 
\end{equation}

where the normalization constant is
\begin{equation}
N=\frac{1}{\left[ \int_0^{\infty} \frac{4 r^2}{a_0^3}\left[ 1-\frac{1}{2}\mu b a_0r^2\right]^2e^{-\frac{2r}{a_0}}dr\right] ^{\frac{1}{2}}}.
\label{psicN}
\end{equation}
where,$\mu=\frac{m_qm_Q}{m_q+m_Q}$ is the reduced mass,$"b"$ is the confinement parameter and $a_0= \frac{3}{4\mu\alpha_s}$. \\\\
Now,while using linear part as perturbation,it was observed [2],[6],that perturbation is valid for very small value of the confinement parameter $b\sim 0.03 GeV^2$. But this value is very small compared to the values obtained from Charmonium spectroscopy ($b\sim 0.183 GeV^2$).To overcome this problem,an additional scaling parameter c was introduced in
Cornell potential,which can accommodate physical value of confinement parameter b .
But,this is in contradiction with the quantum mechanical idea that the presence of a constant term in the potential should not effect the wave-function[6]. Hence,while using Dalgarno,s perturbation theory to heavy flavour physics,we are restricted to only one option with the Cornell Potential that Linear term must be considered as parent.\\\\

$2.$\underline{Linear Parent Coulomb Perturbation}:\\\\
For the second option,we consider  the linear as parent and Coulomb as perturbation and following Dalgarno's perturbation technique we get the total wave function as,
\begin{equation}
\psi^{tot}(r)=\frac{N}{r}[Ai[\varrho]-B(a_1+a_2r+a_3r^2)]
\end{equation}
where,$B=\frac{4\alpha_s}{3}$ ,$\varrho=(2\mu b)^{\frac{1}{3}}r+\varrho_0 $, $\varrho_0$ are the zeroes of Airy Function and,
\begin{eqnarray}
a_1=\frac{0.8808(b\mu)^{\frac{1}{3}}}{E}-\frac{a_3}{\mu E}+\frac{4W'\times 0.21}{3\alpha_s E} \\
a_2=\frac{ba_1}{c}+\frac{4W'\times 0.8808(b\mu)^{\frac{1}{3}}}{3\alpha_s E}-\frac{0.6535\times (b\mu)^{\frac{1}{3}}}{E}\\
a_3=\frac{4\mu W'\times 0.1183}{3\alpha_s}
\end{eqnarray}
With,
\begin{eqnarray}
W^{0}=E=-(\frac{b^2}{2\mu})^{\frac{1}{3}}\varrho_0           \\
W'=4\pi \int_{0}^{\infty} r^2H'\vert \psi(r)\vert ^2dr
\end{eqnarray}

While dealing with the Airy function as the trial wave-function of variational method with Cornell potential,the main problem is that the wave-function has got a singularity at $r=0$. The presence of singularity in a wave-function is not new and in QED also such singularities appear.  
Therefore as discussed in our previous work[18],to calculate the wave-function at the origin,we follow a second method as Quigg[20].  In this method,the wave-function at the origin is found from the condition,$\vert\psi(0)\vert^2=\frac{\mu}{2\pi}\langle\frac{\delta V}{\delta r}\rangle$,and comparing with the works of Quigg,we find  the variational parameter $b'$.

\subsection{Variationally Improved Perturbation Theory:}
In VIPT the total Hamiltonian is expressed as[11-15],
\begin{equation}
H=H_0+H'
\end{equation}
where,$H_0$ is the parent hamiltonian of a parameter $P'$ and $H'$ is the perturbed hamiltonian.  In VIPT,
\begin{equation}
P=P+P'-P'
\end{equation}
$P'$ is the variational method and,
\begin{eqnarray}
H=H_{OP'}+H_0-H_{OP'}+H'\\
=H_{OP'}+H'_{P'}
\end{eqnarray}
The parent hamiltonian $H_{OP'}$ and perturbed hamiltonian $H_{P'}$ now depends on the variational parameter $P'$. Similarly,in the wave-function $P$ will be replaced by $P'$ which is treated as the trial wave-function. Applying,variational method we can calculate the energy eigenvalue and by minimising that the variational parameter is calculated.  
To,this variationally improved wave-function stationary state perturbation theory is applied and the wave-function corrected upto first order is given by,

\begin{equation}
\psi(j)=\psi_{j}^{0}+{\Sigma_{k\neq j}}\frac{\int \psi_k^{0*}H_{P'}\psi_j^{0}dv}{E_j^{0}-E_k^{0}}
\end{equation}
The energy corrected upto first order,
\begin{equation}
E_j=\int \psi_j^{0*}(H_{OP'}+H_P') \psi_j^{0} dv
\end{equation}
where $\psi(k)$ and $E_k$ are the wave function and energy eigenvalues of the $k_th$ state which
are orthonormal to $j_{th}$ state. The superscript $(0)$ is the $zero^{th}$-order correction of
the corresponding quantities.

\subsection{VIPT with Coulomb Parent and Linear perturbation:}
Let us consider a trial wave-function as,
\begin{equation}
\psi_{10}^{(0)}(r)=\frac{(\mu \alpha_{10}')^{\frac{3}{2}}}{\sqrt{\pi}}e^{-\mu \alpha'_{10} r}
\end{equation}
 Where,$\alpha'$ is the variational parameter.  Now following variational method, we get ,
 \begin{equation}
 E(\alpha_{10}')=<\psi\mid H\mid \psi> = \frac{1}{2}\mu {\alpha'_{10}}^2-A\mu \alpha'_{10}+\frac{3b}{2\mu \alpha'_{10}}
 \end{equation}
where,$A=\frac{4\alpha_s}{3}$,and $\alpha_s$ is the strong coupling constant.Now,minimising ,$\frac{dE}{d\alpha'_{10}}=0$ ,we get,
\begin{eqnarray}
{\alpha'_{10}}^3-A{\alpha'_{10}}^2-\frac{3b}{2\mu^2}=0
\end{eqnarray}
This equation is solved by using $Mathematica7$ and we find the variational parameter for different Heavy Flavour mesons which is shown in $Table.1$.

So,we now replace $\alpha'_{10}$ by the variational parameter $\bar{\alpha'}_{10}$.  Now the $k_{th}$ state in the summation of equation $(5)$ which is the 2s state is given by,
\begin{eqnarray}
\psi_{k}^{(0)}(\bar{\alpha'}_{10})=\psi_{20}^{(0)}(\bar{\alpha'}_{10})    \\
                      =\frac{(\mu \bar{\alpha'}_{10})^{\frac{3}{2}}}{\sqrt{8\pi}}(1-\frac{\mu \bar{\alpha'}_{10}r}{2})e^{-\frac{\mu \bar{\alpha'}_{10}r}{2}}
\end{eqnarray}
Hence,the corrected wave-function upto first order is,
\begin{equation}
\psi(\bar{\alpha'}_{(10)})=\psi_{(10)}^{0}(\bar{\alpha'}_{10})+\frac{\int \psi_{20}^{0*}H_{0\bar{\alpha'}_{10}}\psi_{\bar{\alpha'}_{10}}^{0}}{E_{(10)}^{0}-E_{(20)}^{0}}dv\psi_{20}^{0} 
\end{equation}

The summation in equation $(27)$ is dropped since we are considering single $k_{th}$ state.  Now substituting the wave-functions we get the total wave-function corrected upto first order as,

\begin{equation}
\psi(\bar{\alpha}'_{10})=N[\frac{(\mu \bar{\alpha'}_{10})^{\frac{3}{2}}}{\sqrt{\pi}}e^{-\mu \bar{\alpha'}_{10} r}-\frac{4\sqrt{\mu}}{3\sqrt{\pi \bar{\alpha'}_{10}}}(\frac{4\mu\bar{\alpha'}_{10}(\alpha -\bar{\alpha'})}{27}-\frac{32b}{81\mu \bar{\alpha'}_{10}})(1-\frac{\mu \bar{\alpha'}_{10}r}{2})e^{-\frac{\mu \bar{\alpha'}_{10}r}{2}}]
\end{equation}
where "N" is the normalization constant and is obtained from,
\begin{equation}
\int_{0}^{\infty}4\pi r^2 \vert \psi(\bar{\alpha}'_{10}) \vert ^2 dr=1
\end{equation}
\subsection{VIPT with Linear parent and Coulomb perturbation:}
For another analysis we consider the trial wave-function as Airy function,
\begin{equation}
\psi(r)=\frac{1}{2\sqrt{\pi}r}A_{i}[(2 \mu b')r^{\frac{1}{3}}+\varrho _{0n}]
\end{equation}
Here,$b'$ is the variational parameter and $\varrho_{0n}$ are the zeroes of airy function such that $A_{i}[\varrho_{0n}]=0$,and is given as[]:
\begin{equation}
\varrho_{0n}=-[\frac{3\pi(4n-1)}{8}]^{\frac{2}{3}}
\end{equation}
\pagebreak
For different S states few zeroes of the Airy function is listed below:

\begin{tablehere}\scriptsize
\begin{center}
\caption{Zeroes of Airy function for different S-states:}
\begin{tabular}{|c|c|}
 \hline
          
  States&$\varrho_{0n}$                  \\\hline
1s(n=1,l=0)&-2.3194                     \\\hline
2s(n=2,l=0)& -4.083                         \\\hline
3s(n=3,l=0)& -5.5182                      \\\hline
4s(n=4,l=0)&-6.782                           \\\hline
\end{tabular}
\end{center}
\end{tablehere}

It is worthwhile to mention that wave-functions containing Airy function are the solutions of Schrodinger equation for linear confinement potential and it is an infinite series in itself,as:
\begin{equation}
A_i[\varrho]=a_0[1+\frac{\varrho ^3}{3!}+\frac{\varrho ^6}{6!}+\frac{\varrho ^9}{9!}+.....]-b_0[\varrho+\frac{\varrho ^4}{4!}+\frac{\varrho ^7}{7!}+\frac{\varrho ^{10}}{10!}+.....]
\end{equation}
with,$a_0=0.3550281$ and $b_0=0.2588194$.    The corresponding energies are[12],
\begin{equation}
E_n=-[\frac{b'^2}{2\mu}]^{\frac{1}{3}}\varrho_{0n}
\end{equation}

Now we consider the single $k_th$ state in the summation of equation (7) which is the 2s state given by,
\begin{equation}
\psi_{20}^{(0)}=\frac{1}{2\sqrt{\pi}r}Ai[(2\mu b')^{\frac{1}{3}}r-4.083]
\end{equation}
Therefore,the wave-function corrected upto first order following VIPT,
\begin{equation}
\psi_T=N[\psi_{(10)}+\frac{(2\mu)^{\frac{1}{3}}}{(\varrho_{02}-\varrho_{01})b'^{\frac{2}{3}}}((b-b')<r>_{2,1}-\alpha <\frac{1}{r}>_{2,1})\psi_{20}(r)]
\end{equation}
where,
\begin{equation}
<r>_{2,1}=\int_{0}^{\infty} r A_{i}[(2\mu b')^{\frac{1}{3}}r-2.3194]A_{i}[(2\mu b')^{\frac{1}{3}}r-4.083]dr
\end{equation} 
and $\alpha=\frac{4\alpha_s}{3}$,also,
\begin{equation}
<\frac{1}{r}>_{2,1}=\int_{0}^{\infty} \frac{1}{r} A_{i}[(2\mu b')^{\frac{1}{3}}r-2.3194]A_{i}[(2\mu b')^{\frac{1}{3}}r-4.083]dr
\end{equation}
The normalization constant is obtained from,
\begin{equation}
\int_{0}^{\infty}4\pi r^2 \vert \psi_T \vert ^2 dr=1
\end{equation}

\subsection{Mass and Decay constants of Heavy Flavour mesons in VIPT:}
Taking into account the energy shift of mass splitting due to spin interaction in the perturbation theory,the mass formula for pseudo-scalar mesons is given by[21],[22] ,
\pagebreak
\begin{eqnarray}
M_P = m_Q +m_{\overline{Q}}-\frac{8\pi\alpha}{3 m_Qm_{\bar{Q}}}\vert \psi(0) \vert ^2
\end{eqnarray}

For pseudo-scalar mesons,the decay constant $f_p$ is related to the ground state wave function at the origin $\psi(0)$ according to Van-Royen-Weisskopf formula[23], in the non-relativistic limit as,

\begin{equation}
f_p=\sqrt{\frac{12{\mid\psi(0)\mid}^2}{M_p}} 
\end{equation}
where,$M_p$ is the mass of pseudo-scalar meson .  Now with QCD correction factor it can be written as,
\begin{equation}
f_p=\sqrt{\frac{12{\mid\psi(0)\mid}^2}{M_p}{\bar{C}^2}}
\end{equation}

And,
\begin{equation}
\bar{C}^2=1-\frac{\alpha_s}{\pi}[2-\frac{m_Q-m_{\bar{Q}}}{m_Q+m_{\bar{Q}}}ln\frac{m_Q}{m_{\bar{Q}}}]
\end{equation}

The ratios of pseudo-scalar decay constants eg.for $B_S$ and $D_s$ meson can be expressed as,
\begin{equation}
\dfrac{f_{B_s}}{f_{D_S}}=\sqrt{\dfrac{M_{D_S}}{M_{B_S}}}\dfrac{\psi_{B_S}(0)}{\psi_{D_S}(0)}
\end{equation}

\subsection{Mass difference of vector and Pseudo-scalar mesons:}\rm
The mass difference between the Pseudo-scalar and vector meson is given by[24],
\begin{equation}
M_{{(Q\bar{Q})}^*} -M_{(Q\bar{Q})}=\frac{8\pi \alpha}{3m_Qm_{\bar{Q}}}{\mid\psi_{Q\bar{Q}}(0)\mid}^2
\end{equation}
where $m_Q$ is the mass of heavy quark and $m_{\bar{Q}}$ is the mass of antiquark.
This is attributed to the hyperfine interaction and $\alpha=\frac{4\alpha_s}{3}$ where $\alpha_s$ is the strong coupling constant.  

\subsection{Oscillation frequency:}
 The neutral $B_d$ and $B_s$ meson mix with their antiparticles by means of Box diagram and involves exchange of $W$ bosons and $u,c,t$ quarks which leads to oscillation between mass eigenstates[25],[26].  The oscillation is parametrized by mixing mass parameter $\bigtriangleup m$ given by,
 
\begin{equation}
\Delta m_B =\frac{G_F^2m_t^2M_{B_q}f_{B_q}^2}{8\pi}g{(x_t)}\eta _t\mid V_{tq}^*V_{tb}\mid ^2 B
\end{equation}

Where,$\eta _t$ is the gluonic correction to oscillation(=0.55[27]) and $B$ is the bag parameter(=1.34[27]) and the parameter $g(x_t)$ is given as [28],

\begin{equation}
g(x_t)=\frac{1}{4}+\frac{9}{4(1-x_t)}-\frac{3}{2(1-x_t)^2}-\frac{3x_t^2}{2(1-x_t)^3}
\end{equation}
and,$x_t=\frac{m_t^2}{M_W^2}$. From data of Particle data group[29],
$m_t=174GeV$,$M_W=80.403GeV$,$\mid V_{tb}\mid=1$,$\mid V_{td}\mid=0.0074$,$\mid V_{ts}\mid=0.04$.  

\section{Results:}\rm
\subsection{Variational Parameter:}
The variational parameter obtained for the two options is tabulated below(Table2 and 3).
\begin{tablehere}\scriptsize
\begin{center}
\caption{Variational parameter for Coulomb trial wave-function as equation(22):}
\begin{tabular}{|c|c|}
 \hline
          
  Mesons&$\alpha'_{10}$                  \\\hline
$D{(c\overline{u}/\overline{c}d)}$&1.7285               \\\hline
$D{(c\overline{s})}$&1.4642                        \\\hline

$B{(u\overline{b}/d\overline{b})}$ & 1.51164 \\\hline
    $B_s{(s\overline{b})}$&1.230 \\\hline
    $B{(\overline{b}c)}$ &0.6978  \\\hline
\end{tabular}
\end{center}
\end{tablehere}
\begin{tablehere}\scriptsize
\begin{center}
\caption{Variational parameter for Airy trial function as equation(39):}
\begin{tabular}{|c|c|}
 \hline
          
  Mesons&$b'$                 \\\hline
$D{(c\overline{u}/\overline{c}d)}$&2.050             \\\hline
$D{(c\overline{s})}$&1.597         \\\hline

$B{(u\overline{b}/d\overline{b})}$ & 1.7269 \\\hline
    $B_s{(s\overline{b})}$&1.2709 \\\hline
    $B{(\overline{b}c)}$ &0.558  \\\hline
\end{tabular}
\end{center}
\end{tablehere}
\subsection{Mass:}
Following the formalism developed in section.2,we calculate the masses of pseudo-scalar mesons which are shown in Table.4. The input values are, $m_{u/d}=0.336Gev$ ,$m_b=4.95GeV$, $m_c=1.55GeV$, $m_s=0.483GeV$ and $b=0.183GeV^2$,also we take $\alpha_s=0.39$ for C-scale and $\alpha_s=0.22$ for b-scale [29]. The calculated masses of different heavy-light mesons are compared with the recent results from lattice QCD [30] and QCD sum rules[31] and also with recent experimental results[29](Table.4).  It can be easily seen that our results with VIPT FOR option Coulomb parent linear perturbation agrees well with those. Specifically,the mass of $D_s$ meson estimated in VIPT with Coulomb parent Linear perturbation is $1.969 GeV$,which is in excellent agreement with Lattice $(1.969GeV)$,QCD sum rule $(1.97GeV)$ and recent PDG data$(1.968\pm0.0033)$.The pattern is more or less similar in other mesons.\\ \\
We calculate the mass difference between pseudo-scalar and vector mesons and compare with available lattice results and is shown in Table5. From the table,it is clear that while calculating mass difference between PS and vector mesons,Dalgarno's method provide results,which are in well agreement with lattice results.
\pagebreak
\begin{table}\scriptsize
\begin{center}
\caption{Mass of pseudo-scalar mesons$(GeV)$:}
\begin{tabular}{|c|c|c|c|c|c|c|c|}
 \hline
Mesons &VIPT(CP)&VIPT (LP)  & Dalgarno[LP] &Lattice[30]& Q.S.R[31]&Exp Mass[29]     \\\hline                

$D{(c\overline{u}/\overline{c}d)}$ &1.811&1.391 &1.83&1.885&1.87& $1.869\pm0.0016$  \\\hline
$D{(c\overline{s})}$  &  1.969&1.69&1.99&1.969&1.97&$1.968\pm0.0033$ \\\hline
$B{(u\overline{b}/d\overline{b})}$&5.14&5.68&5.27&5.283&5.28&$5.279\pm0.0017$ \\\hline
    $B_s{(s\overline{b})}$&5.35&5.43&5.42&5.366&5.37&$5.366\pm0.0024$\\\hline
    
\end{tabular}
\end{center}
\end{table}

\begin{table}\scriptsize
\begin{center}
\caption{Mass difference of pseudoscalar and vector mesons(GeV):}
\begin{tabular}{|c|c|c|c|c|}
\hline
Mesons &   $M_v-M_p(CP)$&$M_v-M_p(LP)$&Dalgarno(Lp)&$M_v-M_p(Lattice)$[41]  \\\hline
$D{(c\overline{u}/cd)}$    &  0.81&0.66 &0.066 &0.067   \\\hline
$D{(c\overline{s})}$    &0.28 & 0.46& 0.058&0.066 \\\hline
$B{(u\overline{b}/d\overline{b})}$  &0.088& 0.11 &0.11&0.034 \\\hline
$B_s{(s\overline{b})}$& 0.16 &0.081&0.092&0.027 \\\hline
$B{(\overline{b}c)}$   &0.0305 &0.025& 0.32 &  \\\hline
\end{tabular}
\end{center}
\end{table}
\subsection{Decay Constant:}
The decay constants of a few H-L mesons are calculated with VIPT and shown in Table.6.Here too,our results with VIPT(Coulomb parent Linear perturbation)is in well agreement with Lattice results[34,35,36],QCD sum rules[37] and recent PDG data[29].As an illustration,the decay constant of $D$ meson is $0.260 Gev^{-1}$ calculated in VIPT with Coulomb parent Linear perturbation,which is close to lattice results $(0.220Gev^{-1})$,QCD sum rules$(0.206Gev^{-1})$ and recent PDG data$(0.205\pm0.85Gev^{-1})$. The pattern is similar for others. Also the ratio,$\frac{f_B}{f_D}=1.23$ agrees well with lattice results  $\frac{f_B}{f_D}=1.16\pm0.06$.
\begin{table}\scriptsize
\begin{center}
\caption{Decay constants of pseudo-scalar mesons$(GeV^{-1})$:}
\begin{tabular}{|c|c|c|c|c|c|c|}
 \hline
Mesons&VIPT(Coulomb Parent)&VIPT(Linear parent)& Dalgarno[LP]&Q.S.R[37]&Lattice[34-36]&Exp value[29] \\\hline
$D{(c\overline{u}/\overline{c}d)}$&0.260 &0.770 &0.195 &$0.206\pm0.002$ &$0.220\pm0.003$&$0.205\pm0.085\pm0.025$         \\\hline
$D{(c\overline{s})}$& 0.312    &0.678  &0.210&$0.245\pm0.015$&$0.258\pm0.001$&$0.254\pm0.059$ \\\hline
$B{(u\overline{b}/d\overline{b})}$ & 0.321&0.426&0.127&$0.193\pm0.012$&$0.218\pm0.005$&$0.198\pm0.014$\\\hline
$B_s{(s\overline{b})}$&0.410 &0.418&0.146&$0.232\pm0.018$&$0.228\pm0.010$&$0.237\pm0.017$\\\hline
$B{(\overline{b}c)}$ &0.578& 0.362 &0.231&&&0.562\\\hline
\end{tabular}
\end{center}
\end{table}
\subsection{Oscillation Frequency:}
The mixing mass parameter $\Delta m$ ,which is connected to the oscillation of neutral mesons is also calculated and compared with the results of QCD sum rules,lattice QCD and exp.data,which
is shown in table.8. Here too,our predictions with VIPT(Coulomb parent Linear perturbation) is in good
agreement with lattice results[38],QCD sum rule result[39]and exp.data[40],[41].
\begin{table}
\begin{center}
\caption{ Mixing mass parameter($ps^{-1}$):}
\begin{tabular}{|c|c|c|c|c|c|}
  \hline
          
   Meson&$\Delta m_B$(Coulomb parent)&$\Delta m_B$ (Linear Parent)& QCD sum rule&lattice&Exp.value  \\\hline
   $B_D$ &$0.47 $ &0.27&0.48[39]&0.63[38]&0.5[40]                                           \\\hline
   $B_S$ &$15.8$ &9.3&$>14.6$[39]&19.6[38]&17.76[41]                                    \\\hline
\end{tabular}
\end{center}
\end{table}
\pagebreak
\subsection{Comparison between Dalgarno and VIPT:}
Let us now differentiate between Dalgarno's perturbation theory and VIPT in the context of wave-function. We choose charged D meson and neutral $D_s$ meson as a representative case.The wave-functions of $D,D_s$ mesons obtained with both Dalgarno's(Fig1,2,3,4) method and VIPT (Fig5,6,7,8)are plotted for both the options(Linear parent coulomb perturbation and Coulomb parent linear perturbation). 
In Dalgarno's method while using coulomb parent linear perturbation,zero of the wave-function occurs at some finite $"r"$ ,because at this point perturbation breaks down.  This feature is absent in VIPT(Fig5,6,7,8),because VIPT allows both the options.

 At the phenomenological level,within the set of parameters used,the Masses and Decay constants of heavy flavour mesons estimated with VIPT for coulomb parent linear perturbation is in better agreement with the results of lattice Qcd,Qcd sum rules and exp.data,while in case of mass difference between pseudo-scalar and vector mesons,Dalgarno with Linear parent coulomb perturbation is better.\\

\begin{figure}[hbtp]
\begin{center}
{\includegraphics[scale=0.802]{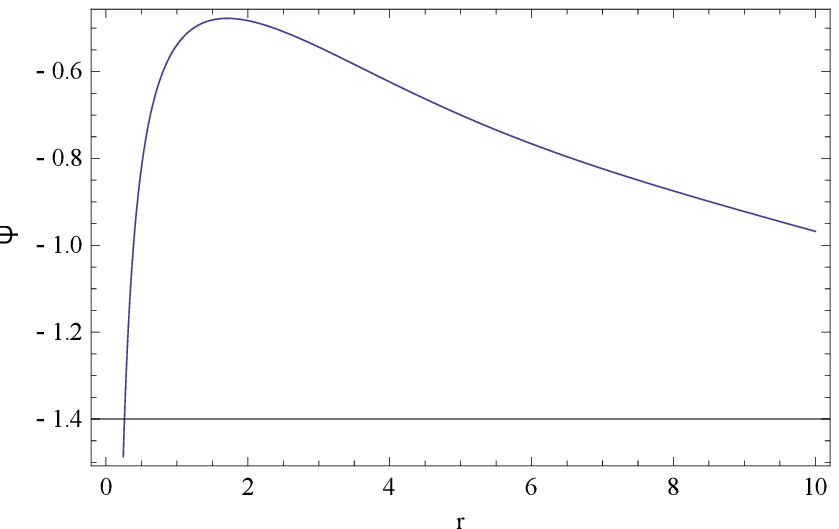}}\quad

\caption{$\psi$ vs r for D meson(Linear parent Coulomb perturbation(Dalgarno))}

{\includegraphics[scale=0.802]{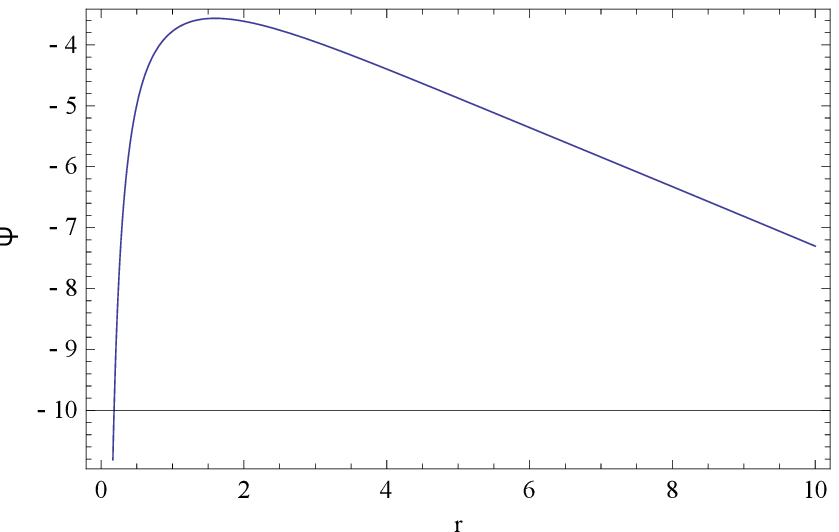}}
\caption{$\psi$ vs r for $D_s$ meson(Linear parent Coulomb perturbation(Dalgarno))}
{\includegraphics[scale=0.802]{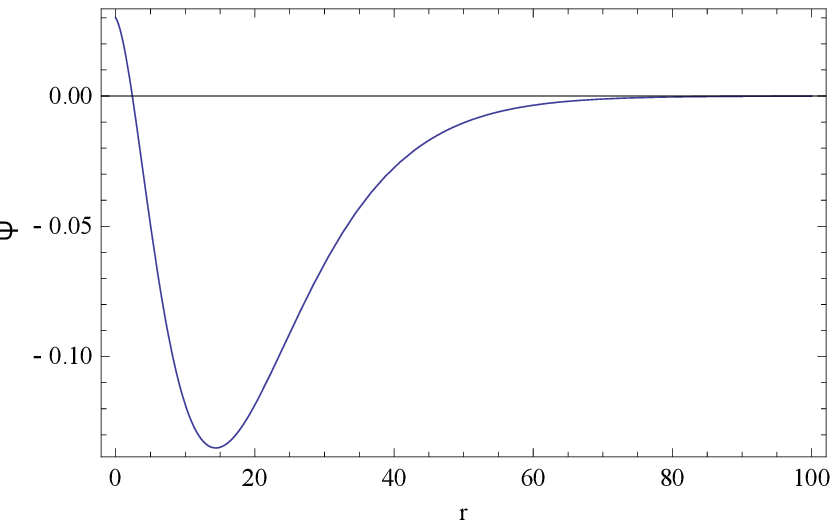}}
\caption{$\psi$ vs r for $D$ meson(Coulomb parent Linear perturbation(Dalgarno))}

{\includegraphics[scale=0.802]{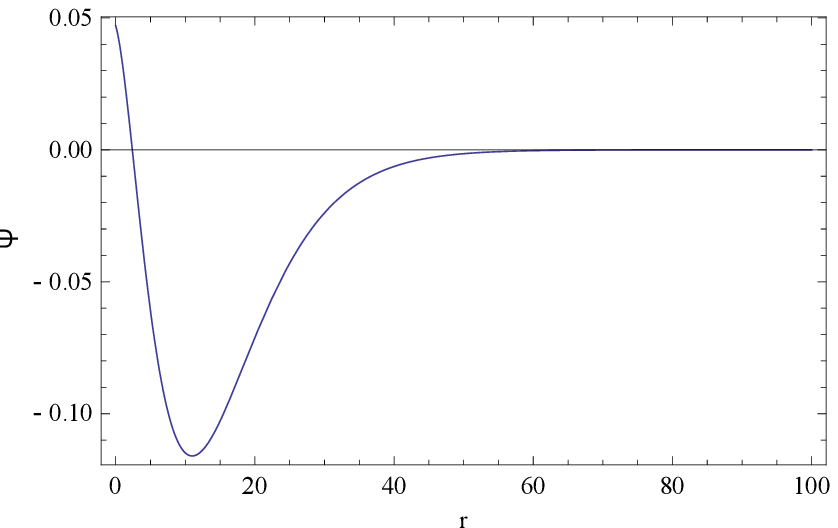}}
\caption{$\psi$ vs r for $D_s$ meson(Coulomb parent Linear perturbation(Dalgarno))}

\label{Fig1}
\end{center}
\end{figure}

\begin{figure}[hbtp]
\begin{center}
{\includegraphics[scale=0.802]{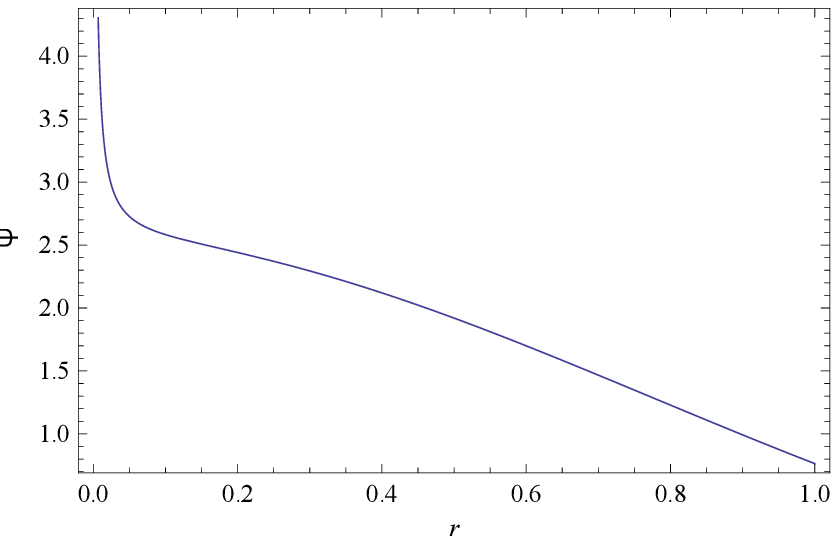}}\quad
\caption{$\psi$ vs r for $D$(Linear parent(VIPT))}
{\includegraphics[scale=0.802]{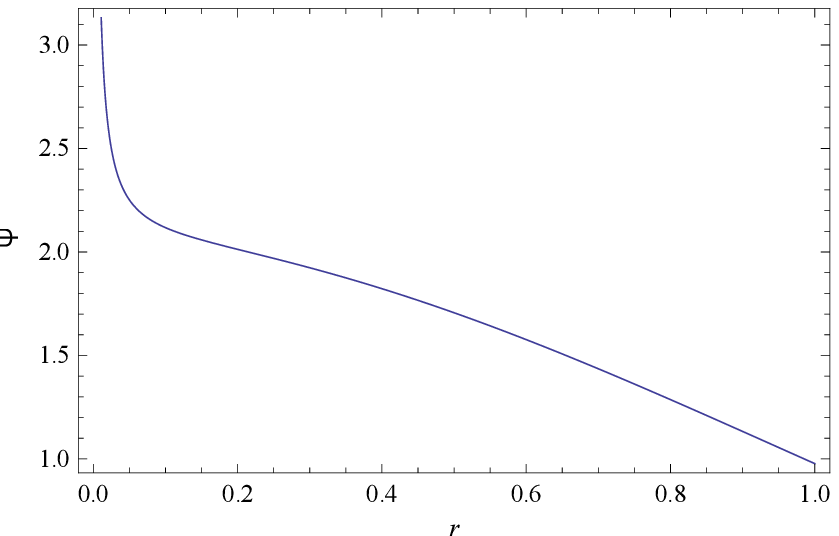}}\quad
\caption{$\psi$ vs r for $D_s$(Linear parent(VIPT))}
{\includegraphics[scale=0.802]{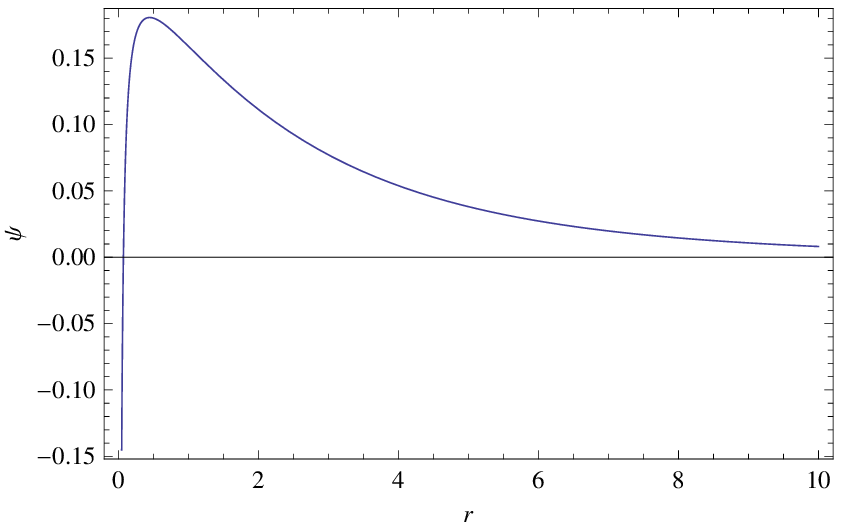}}\quad
\caption{$\psi$ vs r for $D$(Coulomb parent(VIPT))}

{\includegraphics[scale=0.802]{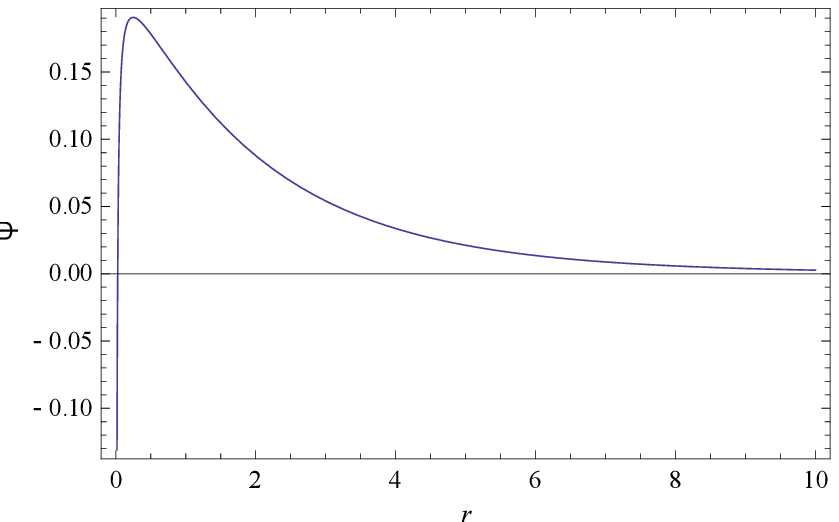}}\quad
\caption{$\psi$ vs r for $D_s$(Coulomb parent(VIPT))}
\label{Fig1}
\end{center}
\end{figure}

\section{Conclusion:}
In this work,we have considered two approximation scheme viz.Dalgarno's perturbation theory and Variationally Improved perturbation theory(VIPT) for Cornell potential to study some properties of Heavy Favour mesons.Detailed comparison is made between the two methods.  It is found that VIPT(coulomb parent linear perturbation) gives good result in masses and decay constants,while in calculating mass difference between pseudo-scalar and vector mesons,Dalgarno(linear parent coulomb perturbation) is better. 
The results of perturbation theory are expressed in terms
of finite power series which converge to the exact values when summation is taken up to higher order. But,in Dalgarno's method, however the results become increasingly worse since the series is
 divergent (being asymptotic). At this juncture, the variational method
which estimates variationally optimized parameters (through energy minimization)
helps in converting the divergent perturbation expansion to a convergent one which
can be evaluated. The variational
method  is quite cumbersome as it is difficult to choose an appropriate trial wave
function in terms of unknown parameter which is later optimized to estimate the
parameter. But in VIPT, we use a known wave function as the trial one (e.g. the
1s state H-atom wave function) and then optimize it to get the new parameter(s)
(e.g. $α10$) which make the perturbation series convergent.

However,sensitivity of our results to the scales of input parameters is a scope of future study.\\
\pagebreak
\section*{Acknowledgement}
 One of the authors (J.L) acknowledges CSIR,India for financial support by providing Fellowship during the research work.


\begin{thebibliography}{99}
\bibitem{Schiff}Quantum Mechanics,L.I.Schiff,Third edition,Mc-Graw Hill Book company,New-york
\bibitem{}D.K.Choudhury and N.S.Bordoloi,Int.J.Mod.Phys.A,Vol. 15, No. 23 (2000) p3667–3678
\bibitem{}N. S. Bordoloi and D. K. Choudhury, Mod. Phys. Lett. A 24, p443 (2009).
\bibitem{kkp}K. K. Pathak and D. K. Choudhury, Chin. Phys. Lett.28, 101201 (2011)
\bibitem{}K.K.Pathak,D.K.Choudhury and N.S.Bordoloi,Int.J.Mod.Phys.A,Vol. 28, No. 2 (2013) 1350010 
\bibitem{T das}T Das and D.K.Choudhury,Int.J of Mod.Phy A,2016,DOI:10.1142/S0217751X1650189X
\bibitem{Hwang}D.S.Hwang etal., Phys. Rev.D 53, 4951,1996
\bibitem{Rai}N.Devlani and A.K.Rai,Phys.Rev.D84,074030,2011
\bibitem{Vega}A.Vega and J.Flores,Pramana – J. Phys. (2016) 87:73
DOI 10.1007/s12043-016-1278-7
\bibitem{}B.J.Hazarika and D.K.Choudhury,Pramana J. Phys.,(2017) 88:56 DOI:10.1007/s12043-016-1357-9
\bibitem{}S K You, K J Jeon, C K Kim and K Nahm, Eur. J. Phys. 19, 179 (1998)
\bibitem{}I J R Aitchison and J J Dudek, Eur. J. Phys. 23, 605 (2002)
\bibitem{}F M Fernandez, Eur. J. Phys. 24, 289 (2003)
\bibitem{}B.J.Hazarika and D.K.Choudhury,Pramana J. Phys.,Vol. 75, No. 3,September 2010,pp. 423–438
\bibitem{}B.J.Hazarika and D.K.Choudhury,Pramana J. Phys.,Vol. 78, No. 4,April 2012,pp. 555–564
\bibitem{}B.J.Hazarika and D.K.Choudhury,Pramana J. Phys.,Vol. 84, No. 1,January 2015,pp. 69–85
\bibitem{}D.K.Choudhury,T.Das and N.S.Bordoli,Arxiv:
\bibitem{}J.Lahkar,D.K.Choudhury and B.J.Hazarika,Commun.theo.Phys.
\bibitem{} N Isgur and M B Wise, Phys. Lett. B232, 113 (1989)
\bibitem{Quigg}Quigg C and Rosner J L 1979 Phys. Rep.56, 167
\bibitem{Griffiths}D Griffiths, Introduction to Elementary Particles;John Wiley and Sons,
New york(1987),p158.
\bibitem{}Halzen and Martin,"Quarks and Leptons",John Wiley and Sons,ISBN:0-471-88741-2,P65
\bibitem{van}Van Royen R et al., Nuovo Cimento \textbf{50}, (1967)
\bibitem{Rai}A.K.Rai,R.H.Parmar and P.C.Vinodkumar,J. Phys. G: Nucl. Part. Phys. 28 (2002) 2275–2282
\bibitem{Buras}Buras A,Phys.Lett.B566,115(2003)DOI:10.1016/S0370-2693(03)00561-6
 \bibitem{}D. Ebert, R. N. Faustov, V. O. Galkin, Phys. Rev. D 67, 014027 (2003).
\bibitem{patel}Bhavin Patel and P C Vinodkumar,Chinese Physics C, Vol. 34, No. 9,(2010); arXiv:hep-ph/0908.2212v1(2009).
\bibitem{}Inami,T and Lim,C.S.Prog.Theo.phys.65(1981);ibid.65(1981)1772(E)297
\bibitem{}C. Patrignani and Particle Data Group, 2016 Chinese Phys. C 40 100
\bibitem{}R.j.Dowdall etal.,HPQCD Collab.,arxiv:1207.5149v1
\bibitem{}Z.G.Wang,Eur. Phys. J. C75 (2015) 427,DOI:10.1140/epjc/s10052-015-3653-9
\bibitem{} K. C. Bowler et al., (UKQCD Collaboration), hep-lat/0007020.
\bibitem{}Heechang Naetal.,Phys. Rev. D 86 (2012) 034506,DOI:10.1103/PhysRevD.86.034506
\bibitem{}W.Chen etal.,TWQCD Collab.,Phy.Lett.B,736(2014),https://doi.org/10.1016/j.physletb.2014.07.025
\bibitem{}W.Lucha etal.,J. Phys. G: Nucl. Part. Phys. 38 (2011) 105002 (17pp)
\bibitem{}A. Bazavov etal.,Phys.Rev.D93, 113016 (2016)
\bibitem{}C.Gay,Ann.Rev.Nucl.Part.Sci.50:577-641,2000,DOI:10.1146/annurev.nucl.50.1.577
\bibitem{}The LHCb collab.,Arxiv:1304.4741v1[hep-ex]
\bibitem{}The LHCb collab.,Eur.J.of Phys.C(2016)76:412













































































\bibitem{Eichten}E.Eichten etal.,Phys.Rev.D21(1980)203


\bibitem{}D. Asner et al. (Heavy Flavor Averaging Group),arXiv:1010.1589.






\end{thebibliography}
\end{document}